\documentclass[prd,aps,twocolumn,showpacs,floats,floatfix]{revtex4}
\usepackage{graphicx}
\usepackage{amsmath}
\usepackage{subfigure}
\usepackage{appendix}
\usepackage[normalem]{ulem}
\usepackage[utf8]{inputenc}
\usepackage{mathrsfs}
\usepackage{latexsym}
\usepackage{amsmath}
\usepackage{graphicx}
\usepackage{subfigure}
\usepackage{dcolumn}
\usepackage{bm}
\usepackage{amssymb}
\usepackage{latexsym}
\usepackage{ulem}
\usepackage[colorlinks,linkcolor=magenta,anchorcolor=cyan,citecolor=blue]{hyperref}
\usepackage{color}
\usepackage{hyperref}
\usepackage[hyphenbreaks]{breakurl}

\def\be{\begin{equation}}
\def\ee{\end{equation}}
\def\ba{\begin{eqnarray}}
\def\ea{\end{eqnarray}}

\def\nn{\nonumber}

\begin{document}
\title{Searching for gravitational wave echoes in GWTC-1 and O3 events}

\author{Yu-Tong Wang$^{1,2,3}$\footnote{wangyutong@ucas.ac.cn}}
\author{Yun-Song Piao$^{1,2,3,4}$\footnote{yspiao@ucas.ac.cn}}

\affiliation{$^1$ School of Fundamental Physics and Mathematical
Sciences, Hangzhou Institute for Advanced Study, UCAS, Hangzhou
310024, China}

\affiliation{$^2$ International Center for Theoretical Physics
Asia-Pacific, Beijing/Hangzhou, China}

\affiliation{$^3$ School of Physics, University of Chinese Academy
of Sciences, Beijing 100049, China}

\affiliation{$^4$ Institute of Theoretical Physics, Chinese
Academy of Sciences, P.O. Box 2735, Beijing 100190, China}

\begin{abstract}

Gravitational wave (GW) echoes, if they exist, would be a probe to
the near-horizon physics of black hole. In this brief report, we
performed the Monte Carlo Markov Chain analysis to search for echo
signal in all GWTC-1 and O3 GW events. We focus on the
Insprial-Merger-Ringdown-Echo (IMRE) waveform, and apply the
Bayesian model selection to compare the IMRE result with IMR's (no
echo). We find no statistically significant ($<1\sigma$ combined)
evidence for the GW echoes and only individual GW events with the
echoes at $1\sim 2\sigma$ significance.

\end{abstract}

\maketitle

\section{introduction}
Over the past several years, the network of Advanced
LIGO~\cite{TheLIGOScientific:2014jea} and
Virgo~\cite{TheVirgo:2014hva} observatories have demonstrated that
the detection of gravitational wave (GW) has become a significant
method to observe the universe. LIGO and Virgo collaboration have
reported 15 GW events till
now~\cite{LIGOScientific:2018mvr,TheLIGOScientific:2017qsa,LIGOScientific:2020stg,Abbott:2020uma,Abbott:2020tfl,Abbott:2020khf}.

The corresponding GW events are compatible with the binary black
holes (BHs) coalescences predicted by General Relativity (GR)
~\cite{TheLIGOScientific:2016pea,Abbott:2017vtc,TheLIGOScientific:2016src,Abbott:2018lct}.
However, it is still possible that the new physics might emerge
near the horizon e.g.\cite{Giddings:2016tla,Giddings:2017jts}. The
GW signal of binary BHs coalescences consists of the inspiral
phase, the merger phase and the ringdown phase, see
\cite{Cardoso:2017cqb,Cardoso:2019rvt} for a review. It has been
showed in
Refs.\cite{Cardoso:2016rao,Cardoso:2016oxy,Abedi:2016hgu} that if
certain physics (surface or barrier) near horizon reflects GW, the
ringdown waveform of post-merger object will show itself a series
of ``echoes" at late-time. The GW echoes, regarded as the probes
of new physics at the near-horizon regime, have motivated the
searching for the corresponding signals in GW data
\cite{Abedi:2016hgu,Westerweck:2017hus}, see also
\cite{Wang:2019szm,Lo:2018sep,Nielsen:2018lkf,Uchikata:2019frs}.

The property of echo encodes the distinct physics of post-merger
BH or compact object
\cite{Mark:2017dnq,Maselli:2017tfq,Wang:2018mlp,Correia:2018apm,Wang:2018cum,Pani:2018flj,Testa:2018bzd,Konoplya:2018yrp,Ghersi:2019trn,Li:2019kwa,Maggio:2019zyv,Huang:2019veb,Hui:2019aox,Bronnikov:2019sbx,Liu:2020qia,Maggio:2020jml,Dimitrov:2020txx}.
The echo interval might be not constant. It has been pointed out
that if the post-merger object is a wormhole that is slowly
pinching off and eventually collapsing into a BH, the ringdown
waveform will exhibit a series of echoes with increasing intervals
\cite{Wang:2018mlp}, see also primordial compact object
\cite{Wang:2018cum}. In particular, if the near-horizon regime of
BH is modelled as a multiple-barriers filter, the mixing of
echoes, even the superpositions, will be also inevitably present
\cite{Li:2019kwa}. These studies not only enriched the echo
phenomenology, but also helped to the searching for the echo
signals \cite{Wang:2019szm}.

In this brief report, we will perform the MCMC analysis to search
for GW echo signal in all GWTC-1 and O3 events. We focus on the
full Inspiral-Merger-Ringdown-Echo (IMRE) waveform, and apply the
Bayesian model selection to compare the IMRE result with IMR's (no
echo). In addition, we also consider the possibility of the
unequal interval echo \cite{Wang:2018mlp,Wang:2019szm}.

\section{Method and Results}

Here, it is sufficient to consider a simple phenomenological
waveform \ba \Psi_{IMRE}(t) & =& \Psi_{IMR}(t)+\Psi^{echo}(t).
\label{template} \ea where \ba
&\Psi^{echo}&(t)=\sum_{n=1}(-1)^{n}{ A}_{n}e^{-\frac{x^2_{n}}{2
\beta^2}}\nn\\ &&\cos\left[2\pi {f}_n \left[t-t_{echo}-\left(n
+\frac{n(n+1)}{2}r\right)\Delta t_{echo}\right]\right],\nn\ea and
$\Psi_{IMR}(t)$ corresponds to the IMR (each GW event have
different IMR waveform).

We, for simplicity, set ${A}_n\sim {{\cal A}A}/(3+n)$ with ${\cal
A}$ being the ratio of the first echo amplitude to the ringdown
peak $A$ and the increment of echo intervals monotonical and
proportional to $\Delta t_{echo}$: $\delta t=r \Delta t_{echo}$.
As suggested in Refs.\cite{Wang:2018mlp,Wang:2018cum}, $r\neq 0$
will be a hint of wormhole or specific cosmological scenarios.

\begin{figure*}[h!]
\includegraphics[width=19cm,trim={0 0cm 0 0},clip]{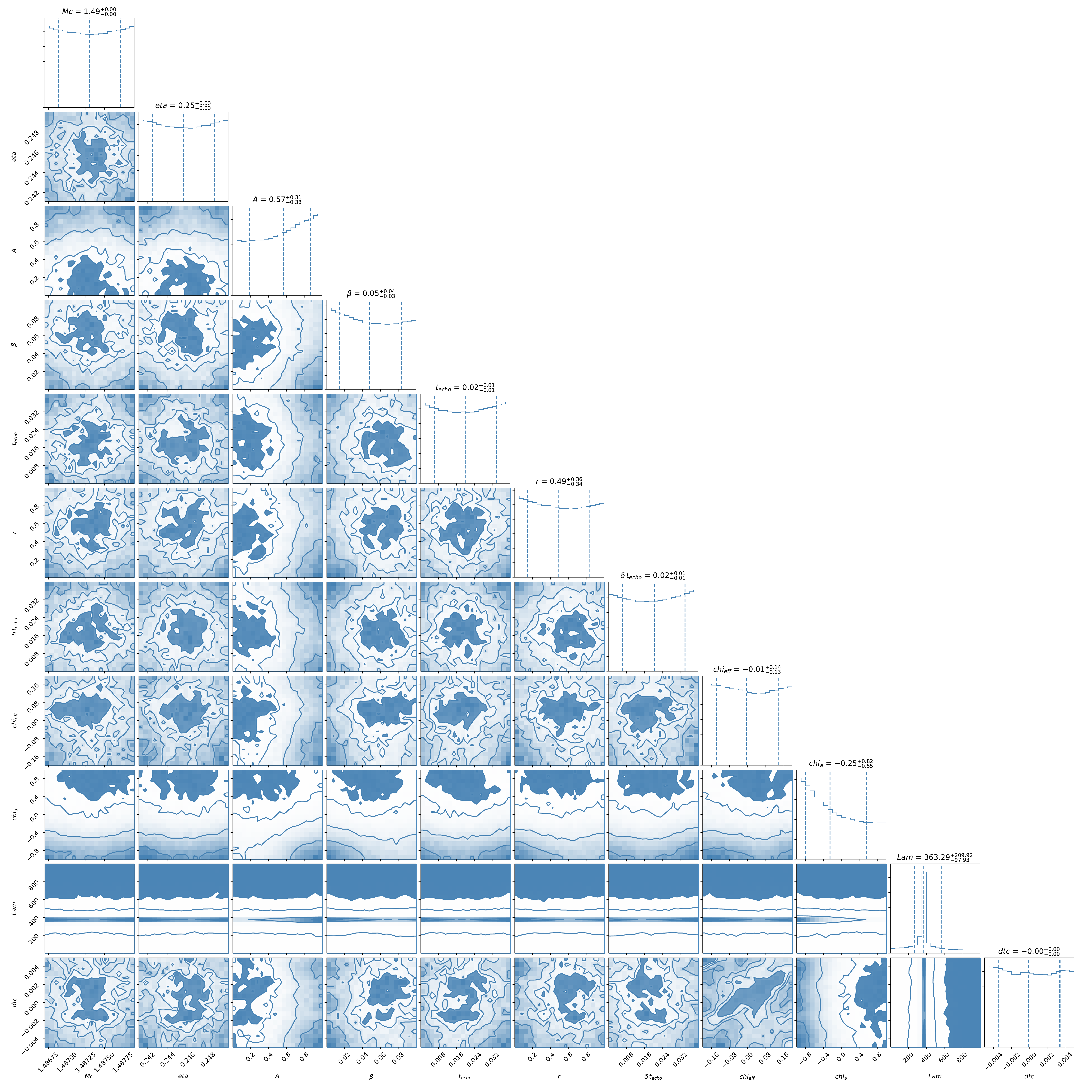}
\caption{Posterior distributions for GW190425, with contours
corresponding to the $68\%$ and $95\%$ regions. }
\label{fig:GW190425}
\end{figure*}

We performed the MCMC analysis with the $\textbf{MGWB}$ (Modified
GWBinning code package)\cite{MGWB} and {\it emcee3.0.2} package
\cite{ForemanMackey:2012ig} on the parameter set \be \{\left({\rm
IMR}\,{\rm parameters}\right), {\cal{A}}, \beta, r,
t_{echo},\Delta t_{echo}\}\label{parameter}\ee of (\ref{template})
for all GWTC-1 and O3 events, see e.g.Fig.\ref{fig:GW190425} for
the GW190425. However, the corresponding results should be
assessed with the Bayesian model selection.

Regarding the IMR and IMRE as hypothesis ${\cal H}_0$ and ${\cal
H}_1$, respectively, we have the logarithm of Bayes factor $\ln
B_{01} = \ln{p(d|{\cal H}_1,I)\over  p(d|{\cal H}_0,I)}$, which
reflects the preference of IMRE compared with IMR. Thus for a set
of independent events with data denoted by $d= \left\{d_1, d_2,
..., d_N\right\}$, we have
\begin{align} \ln{{B}_{0}^{1}} = \ln{\prod_{i = 1}^{N}
\frac{p(d_i|\mathcal{H}_{1}, I)}{p(d_i|\mathcal{H}_{0}, I)}}
 = \sum_{i = 1}^{N}\ln{{{B}_{0i}^{1}}},
\label{B}\end{align} where ${{{B}_{0i}^{1}}}$ corresponds to the
$i$th candidate of GW echoes event.

Tabs.~\ref{tab:GWTC} and \ref{tab:O3} show the Bayes results for
all GWTC-1 and O3 events.

\begin{figure}[!ht]
\begin{center}
\includegraphics[width=1.0\columnwidth]{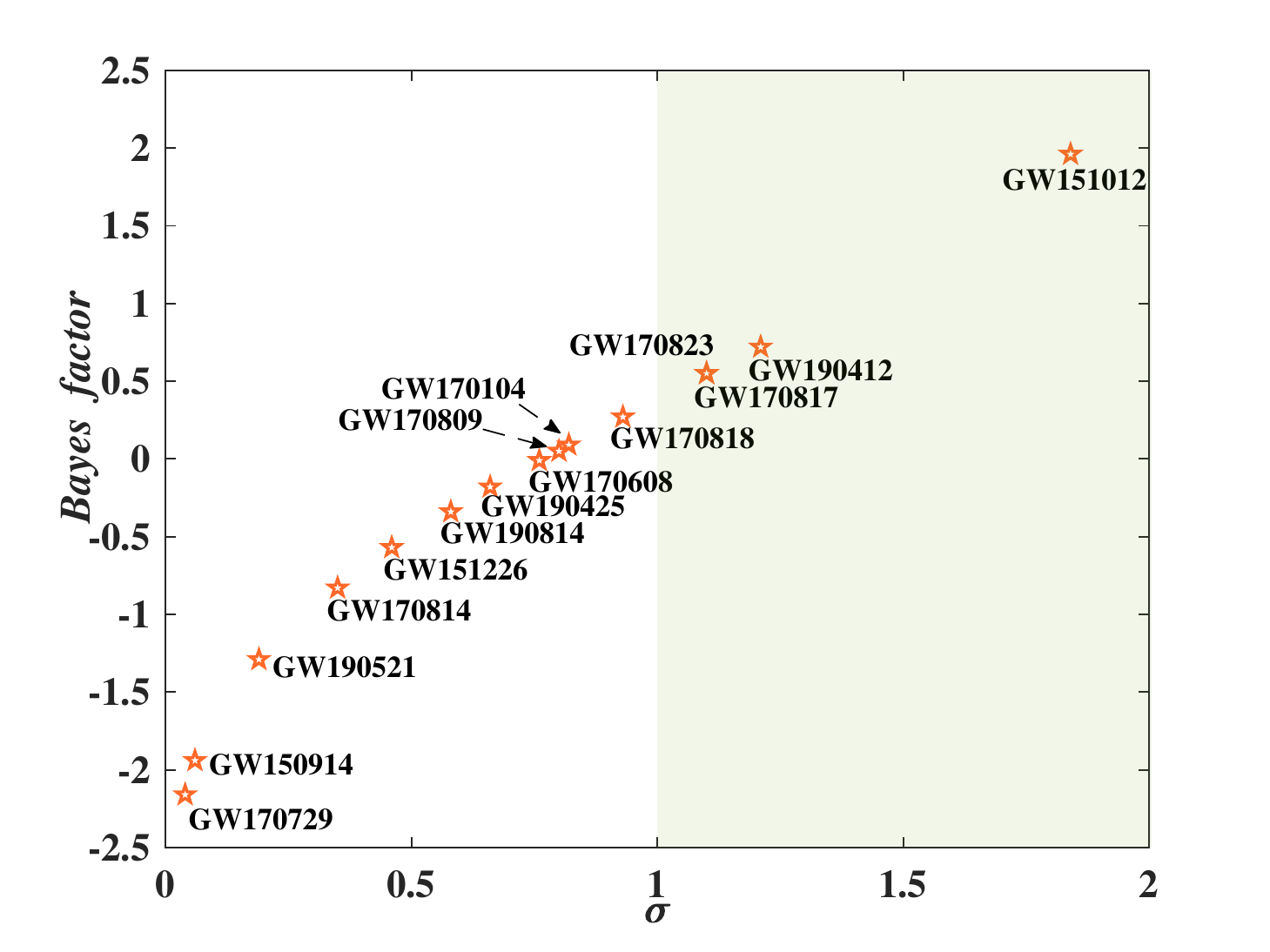}
\caption{The Bayes factor and the statistical significance for
each event. Here, GW170817 and GW170823 have the similar Bayes
factor. }
 \label{fig:BPS}
 \end{center}
\end{figure}

The combined Bayesian factor for GWTC-1 is -1.49, which
corresponding 0.14$\sigma$, while the combined Bayesian factor for
O3 events is -1.08, which is 0.26$\sigma$, larger than that for
GWTC-1. This might be a reflection of slightly higher sensitivity
of Advance LIGO O3. We also show the statistical significance of
each GW event in Fig.\ref{fig:BPS}. Though the log Bayes factor
has no preference for IMRE (or the echo signal), the individual GW
events with $>1\sigma$ significance (1.1$\sigma$ for GW170817 and
GW170823, 1.21$\sigma$ for GW190412, and 1.84$\sigma$ for
GW151012) seem bring out the positive side, which are worth
investigating deeply.

\begin{table}[]
\begin{tabular}{|c|c|c|p{3.3cm}<{\centering}|}
\hline Event & $\ln B$ & $\sigma$ & $\Psi_{IMR}$  \\ \hline
GW150914 & -1.94 & 0.06 & IMRPhenomD \\ \hline GW151012 & 1.96 &
1.84 & IMRPhenomD\\ \hline GW151226 & -0.57 & 0.46 & IMRPhenomD\\
\hline GW170104 & 0.09 & 0.82 & IMRPhenomD \\ \hline GW170608 &
-0.01 &  0.76 & IMRPhenomD\\ \hline GW170729 & -2.16 & 0.04 &
SEOBNRv4\\ \hline GW170809 & 0.05 & 0.80 & IMRPhenomD\\ \hline
GW170814 & -0.83 & 0.35 & IMRPhenomD\\ \hline GW170817 & 0.55 &
1.10 & IMRPhenomDNRTidal\\ \hline GW170818 & 0.27 & 0.93 &
IMRPhenomD\\ \hline GW170823 & 0.55 & 1.10 & IMRPhenomD \\ \hline
Combined & -1.49 & 0.14 &\\ \hline
\end{tabular}
\caption{Log Bayes factors and statistical significances of IMRE
compared with IMR for the GWTC-1 events.} \label{tab:GWTC}
\end{table}

\begin{table}[]
\begin{tabular}{|c|c|c|p{3.3cm}<{\centering}|}
\hline Event  & $\ln B$  & $\sigma$ &  $\Psi_{IMR}$ \\ \hline
GW190814 & -0.34 &  0.58  & IMRPhenomD \\ \hline GW190521 & -1.29
& 0.19 & NRSur7dq4\\ \hline GW190425 & -0.18 & 0.6 & IMRPhenomDNRTidal\\
\hline GW190412 & 0.72 & 1.21 & IMRPhenomPv3HM\\ \hline Combined &
-1.08 & 0.26 & \\ \hline
\end{tabular}
\caption{Log Bayes factors and statistical significance of IMRE
compared with IMR for the O3 GW events.} \label{tab:O3}
\end{table}

\section{Discussion}

We reported the results of searching for GW echo in all GWTC-1 and
O3 events, see Tabs.~\ref{tab:GWTC} and \ref{tab:O3}, and
Fig.\ref{fig:BPS}. We apply the Bayesian model selection to, for
the first time, compare the IMRE result with IMR's (no echo).
Though the echo waveform we consider is quite simplified, our
method is actually independent of the waveform model used, and can
be applicable for other echo waveforms motivated by BH physics.

Through the Bayesian model selection, we found that all GW events
reported in GWTC-1 and O3 so far have no preference for the IMRE,
which suggests no statistically significant (only $<1\sigma$
combined) evidence for the GW echoes. However, individual GW
events seems have the slightly positive results, so it might be
expected that with higher sensitivity of aLIGO O5 or ET, the GW
echo signal would be detectable.

\section{Acknowledgments}

YTW thanks Jun Zhang, Xiaokang Zhou for useful discussions and
helps, and the talk invitation of summer school organized by Bin
Wang and Jian-Pin Wu in Yangzhou University. YTW is supported by
NSFC, No.11805207, the sixty-second batch of China Postdoctoral
Fund. YSP is supported by NSFC, Nos.11690021, 12075246. Our all
computations are performed on the TianHe-II supercomputer.

\begin{thebibliography}{99}

\bibitem{TheLIGOScientific:2014jea}
J.~Aasi \textit{et al.} [LIGO Scientific],
Class. Quant. Grav. \textbf{32}, 074001 (2015)
[arXiv:1411.4547 [gr-qc]].

\bibitem{TheVirgo:2014hva}
F.~Acernese \textit{et al.} [VIRGO],
Class. Quant. Grav. \textbf{32}, no.2, 024001 (2015)
[arXiv:1408.3978 [gr-qc]].

\bibitem{LIGOScientific:2018mvr}
B.~P.~Abbott \textit{et al.} [LIGO Scientific and Virgo],
Phys. Rev. X \textbf{9}, no.3, 031040 (2019)
[arXiv:1811.12907 [astro-ph.HE]].

\bibitem{TheLIGOScientific:2017qsa}
B.~P.~Abbott \textit{et al.} [LIGO Scientific and Virgo],
Phys. Rev. Lett. \textbf{119}, no.16, 161101 (2017)
[arXiv:1710.05832 [gr-qc]].

\bibitem{LIGOScientific:2020stg}
R.~Abbott \textit{et al.} [LIGO Scientific and Virgo],
Phys. Rev. D \textbf{102}, no.4, 043015 (2020)
[arXiv:2004.08342 [astro-ph.HE]].

\bibitem{Abbott:2020uma}
B.~P.~Abbott \textit{et al.} [LIGO Scientific and Virgo],
Astrophys. J. Lett. \textbf{892}, no.1, L3 (2020)
[arXiv:2001.01761 [astro-ph.HE]].

\bibitem{Abbott:2020tfl}
R.~Abbott \textit{et al.} [LIGO Scientific and Virgo],
Phys. Rev. Lett. \textbf{125}, no.10, 101102 (2020)
[arXiv:2009.01075 [gr-qc]].

\bibitem{Abbott:2020khf}
R.~Abbott \textit{et al.} [LIGO Scientific and Virgo],
Astrophys. J. Lett. \textbf{896}, no.2, L44 (2020)
[arXiv:2006.12611 [astro-ph.HE]].

\bibitem{TheLIGOScientific:2016pea}
B.~P.~Abbott \textit{et al.} [LIGO Scientific and Virgo],
Phys. Rev. X \textbf{6}, no.4, 041015 (2016)
[erratum: Phys. Rev. X \textbf{8}, no.3, 039903 (2018)]
[arXiv:1606.04856 [gr-qc]].

\bibitem{Abbott:2017vtc}
B.~P.~Abbott \textit{et al.} [LIGO Scientific and VIRGO],
Phys. Rev. Lett. \textbf{118}, no.22, 221101 (2017)
[erratum: Phys. Rev. Lett. \textbf{121}, no.12, 129901 (2018)]
[arXiv:1706.01812 [gr-qc]].

\bibitem{TheLIGOScientific:2016src}
B.~P.~Abbott \textit{et al.} [LIGO Scientific and Virgo],
Phys. Rev. Lett. \textbf{116}, no.22, 221101 (2016)
[erratum: Phys. Rev. Lett. \textbf{121}, no.12, 129902 (2018)]
[arXiv:1602.03841 [gr-qc]].

\bibitem{Abbott:2018lct}
B.~P.~Abbott \textit{et al.} [LIGO Scientific and Virgo],
Phys. Rev. Lett. \textbf{123}, no.1, 011102 (2019)
[arXiv:1811.00364 [gr-qc]].

\bibitem{Giddings:2016tla}
  S.~B.~Giddings,
  Class.\ Quant.\ Grav.\  {\bf 33}, no. 23, 235010 (2016)
  [arXiv:1602.03622 [gr-qc]].

\bibitem{Giddings:2017jts}
  S.~B.~Giddings,
  Nature Astronomy 1, Article number: 0067 (2017)
  [arXiv:1703.03387 [gr-qc]].

\bibitem{Cardoso:2017cqb}
V.~Cardoso and P.~Pani,
Nature Astron. \textbf{1}, no.9, 586-591 (2017)
[arXiv:1709.01525 [gr-qc]].

\bibitem{Cardoso:2019rvt}
  V.~Cardoso and P.~Pani,
  Living Rev.\ Rel.\  {\bf 22}, no. 1, 4 (2019)
  [arXiv:1904.05363 [gr-qc]].

\bibitem{Cardoso:2016rao}
V.~Cardoso, E.~Franzin and P.~Pani,
Phys. Rev. Lett. \textbf{116}, no.17, 171101 (2016)
[erratum: Phys. Rev. Lett. \textbf{117}, no.8, 089902 (2016)]
[arXiv:1602.07309 [gr-qc]].

\bibitem{Cardoso:2016oxy}
V.~Cardoso, S.~Hopper, C.~F.~B.~Macedo, C.~Palenzuela and P.~Pani,
Phys. Rev. D \textbf{94}, no.8, 084031 (2016)
[arXiv:1608.08637 [gr-qc]].

\bibitem{Abedi:2016hgu}
J.~Abedi, H.~Dykaar and N.~Afshordi,
Phys. Rev. D \textbf{96}, no.8, 082004 (2017)
[arXiv:1612.00266 [gr-qc]].

\bibitem{Westerweck:2017hus}
J.~Westerweck, A.~Nielsen, O.~Fischer-Birnholtz, M.~Cabero, C.~Capano, T.~Dent, B.~Krishnan, G.~Meadors and A.~H.~Nitz,
Phys. Rev. D \textbf{97}, no.12, 124037 (2018)
[arXiv:1712.09966 [gr-qc]].

\bibitem{Wang:2019szm}
Y.~T.~Wang, J.~Zhang, S.~Y.~Zhou and Y.~S.~Piao,
Eur. Phys. J. C \textbf{79}, no.9, 726 (2019)
[arXiv:1904.00212 [gr-qc]].

\bibitem{Lo:2018sep}
R.~K.~L.~Lo, T.~G.~F.~Li and A.~J.~Weinstein,
Phys. Rev. D \textbf{99}, no.8, 084052 (2019)
[arXiv:1811.07431 [gr-qc]].

\bibitem{Nielsen:2018lkf}
A.~B.~Nielsen, C.~D.~Capano, O.~Birnholtz and J.~Westerweck,
Phys. Rev. D \textbf{99}, no.10, 104012 (2019)
[arXiv:1811.04904 [gr-qc]].

\bibitem{Uchikata:2019frs}
N.~Uchikata, H.~Nakano, T.~Narikawa, N.~Sago, H.~Tagoshi and T.~Tanaka,
Phys. Rev. D \textbf{100}, no.6, 062006 (2019)
[arXiv:1906.00838 [gr-qc]].

\bibitem{Mark:2017dnq}
Z.~Mark, A.~Zimmerman, S.~M.~Du and Y.~Chen,
Phys. Rev. D \textbf{96}, no.8, 084002 (2017)
[arXiv:1706.06155 [gr-qc]].

\bibitem{Maselli:2017tfq}
  A.~Maselli, S.~H.~V\"{o}lkel and K.~D.~Kokkotas,
  Phys.\ Rev.\ D {\bf 96}, no. 6, 064045 (2017)
  [arXiv:1708.02217 [gr-qc]].

\bibitem{Wang:2018mlp}
Y.~T.~Wang, Z.~P.~Li, J.~Zhang, S.~Y.~Zhou and Y.~S.~Piao,
Eur. Phys. J. C \textbf{78}, no.6, 482 (2018)
[arXiv:1802.02003 [gr-qc]].





\bibitem{Correia:2018apm}
M.~R.~Correia and V.~Cardoso,
Phys. Rev. D \textbf{97}, no.8, 084030 (2018)
[arXiv:1802.07735 [gr-qc]].

\bibitem{Wang:2018cum}
Y.~T.~Wang, J.~Zhang and Y.~S.~Piao,
Phys. Lett. B \textbf{795}, 314-318 (2019)
[arXiv:1810.04885 [gr-qc]].

\bibitem{Pani:2018flj}
P.~Pani and V.~Ferrari,
Class. Quant. Grav. \textbf{35}, no.15, 15LT01 (2018)
[arXiv:1804.01444 [gr-qc]].

\bibitem{Testa:2018bzd}
A.~Testa and P.~Pani,
Phys. Rev. D \textbf{98}, no.4, 044018 (2018)
[arXiv:1806.04253 [gr-qc]].

\bibitem{Konoplya:2018yrp}
R.~A.~Konoplya, Z.~Stuchl\'\i{}k and A.~Zhidenko,
Phys. Rev. D \textbf{99}, no.2, 024007 (2019)
[arXiv:1810.01295 [gr-qc]].

\bibitem{Ghersi:2019trn}
J.~T.~G\'alvez Ghersi, A.~V.~Frolov and D.~A.~Dobre,
Class. Quant. Grav. \textbf{36}, no.13, 135006 (2019)
[arXiv:1901.06625 [gr-qc]].

\bibitem{Li:2019kwa}
Z.~P.~Li and Y.~S.~Piao,
Phys. Rev. D \textbf{100}, no.4, 044023 (2019)
[arXiv:1904.05652 [gr-qc]].

\bibitem{Maggio:2019zyv}
E.~Maggio, A.~Testa, S.~Bhagwat and P.~Pani,
Phys. Rev. D \textbf{100}, no.6, 064056 (2019)
[arXiv:1907.03091 [gr-qc]].

\bibitem{Huang:2019veb}
Y.~X.~Huang, J.~C.~Xu and S.~Y.~Zhou,
Phys. Rev. D \textbf{101}, no.2, 024045 (2020)
[arXiv:1908.00189 [gr-qc]].

\bibitem{Hui:2019aox}
  L.~Hui, D.~Kabat and S.~S.~C.~Wong,
  JCAP {\bf 1912}, 020 (2019)
  [arXiv:1909.10382 [gr-qc]].


\bibitem{Bronnikov:2019sbx}
K.~A.~Bronnikov and R.~A.~Konoplya,
Phys. Rev. D \textbf{101}, no.6, 064004 (2020)
[arXiv:1912.05315 [gr-qc]].

\bibitem{Liu:2020qia}
H.~Liu, P.~Liu, Y.~Liu, B.~Wang and J.~P.~Wu,
[arXiv:2007.09078 [gr-qc]].

\bibitem{Maggio:2020jml}
E.~Maggio, L.~Buoninfante, A.~Mazumdar and P.~Pani,
Phys. Rev. D \textbf{102}, no.6, 064053 (2020)
[arXiv:2006.14628 [gr-qc]].

\bibitem{Dimitrov:2020txx}
V.~Dimitrov, T.~Lemmens, D.~R.~Mayerson, V.~S.~Min and B.~Vercnocke,
[arXiv:2007.01879 [hep-th]].

\bibitem{MGWB}
\href{https://bitbucket.org/YutongWang/workspace/projects/MGWB}{https://bitbucket.org/YutongWang/MGWB}

\bibitem{ForemanMackey:2012ig}
D.~Foreman-Mackey, D.~W.~Hogg, D.~Lang and J.~Goodman,
Publ. Astron. Soc. Pac. \textbf{125}, 306-312 (2013)
[arXiv:1202.3665 [astro-ph.IM]].





\end{thebibliography}
\end{document}